\documentclass[prc,showpacs,floatfix]{revtex4}
\usepackage{bm}
\usepackage{dcolumn}
\usepackage{graphicx}
\usepackage{amsmath}
\usepackage{amssymb}
\usepackage{color}
\usepackage{bbold}
\newcommand{\beq}{\begin{equation}}
\newcommand{\eeq}{\end{equation}}
\newcommand{\beqa}{\begin{eqnarray}}
\newcommand{\eeqa}{\end{eqnarray}}
\definecolor{green2}{rgb}{0,0.6,0.1}

\begin{document}
\title{
\hfill{\small {\bf MKPH-T-07-02}}\\
{\bf Inclusive electron scattering off $^4$He}}
\author{Sonia Bacca$^{a}$\footnote{electronic address: s.bacca@gsi.de},
 Hartmuth Arenh{\"o}vel$^{b}$, Nir Barnea$^{c}$, Winfried Leidemann$^{d}$
and Giuseppina Orlandini$^{d}$}

\affiliation{
$^{a}$Gesellschaft f\"{u}r Schwerionenforschung, Planckstr.~1,
D-64291 Darmstadt, Germany \\
$^{b}$Institut f\"ur Kernphysik, Johannes Gutenberg-Universit\"at,
J.J.-Becher-Weg~45, D-55099 Mainz, Germany\\
$^{c}$Racah Institute of Physics, Hebrew University, 91904, Jerusalem,
Israel\\ 
$^{d}$Dipartimento di Fisica, Universit\`{a} di Trento and INFN\\
(Gruppo Collegato di Trento), via Sommarive 14, I-38100 Trento, Italy}
\date{\today}
\begin{abstract}
Inclusive electron scattering off $^4$He is
investigated for low and medium energy and momentum transfers. The final
state interaction, given by the simple semirealistic Malfliet-Tjon
potential, is treated rigorously applying the Lorentz Integral
Transform (LIT) method. Besides the nonrelativistic one-body current a
consistent meson exchange current is constructed and
implemented. Results are presented for both longitudinal and
transverse response functions at various momentum transfers. Good
agreement with experimental data is found for the longitudinal
response function, while some strength is missing in the transverse
response function on the low-energy side of the quasi-elastic peak. 
\end{abstract}

\maketitle
\section{Introduction}
In recent years considerable progress has been achieved in rigorous
microscopic nuclear structure calculations of few-body nuclei with
realistic or at least semirealistic interactions. Quite different
approaches have been devised for bound state problems 
(see~\cite{bench_he4}).
For break-up observables the powerful method of the ``Lorentz
Integral Transform'' (LIT) has  been developed  in  Ref.~\cite{EfL94}. The
method is very convenient for calculating cross sections of perturbation
induced reactions, where the inelastic response of the nuclear system is given by an absolute square of a transition matrix element summed over
the whole spectrum of final states. Within this approach
one does not evaluate the response function straightforwardly, but
calculates instead an integral transform of the response with a
Lorentzian kernel. The response function itself is then obtained by
inverting the transform. The specific advantage of this method lies in
the fact that no explicit calculation of the complicated spectrum of
final state wave functions is required. 

Over recent years, the LIT method has been exploited very
successfully in the calculation of electromagnetic
inclusive responses of various light nuclei (see,
e.g.~\cite{BaM01,BaA04,GaB06}), of neutrino inelastic reactions
\cite{GaB04,GaB07}, and even of exclusive processes (see,
e.g.~\cite{QuL04,QuE05}). With respect to inclusive
electron scattering considered here, up to now only the longitudinal
response function $R_L$, induced by the electromagnetic charge
${\rho}({\boldsymbol q})$, has been investigated in the LIT approach, namely
for $^3$H and $^3$He with realistic nuclear forces~\cite{EfL04,EfL05}
and for $^4$He with the semirealistic TN potential~\cite{EfL97,EfL98}, 
while the transverse response $R_T$, induced by the current density operator
${\boldsymbol j}({\boldsymbol q})$, has not yet been considered within this method. 

In the present work we evaluate both inclusive responses of the
four-body system $^4$He. With the exception of Refs.~\cite{EfL97,EfL98},
most previous calculations of $R_L$ and $R_T$ relied on the plane wave 
impulse approximation (PWIA) (see, e.g.~\cite{pwia}), where the
interaction among the final reaction products is neglected (PW) and
where only one-body electromagnetic operators are considered (IA).
In Ref.~\cite{CaS92,CaS94} the  Laplace transforms of the two responses
(the so called ``Euclidean responses'') were calculated using the
Green Function Monte Carlo (GFMC) method with a realistic interaction
and a consistent treatment of additional meson exchange currents (MEC).
This work showed that i) the final state interaction (FSI) is relevant
in light nuclei, especially in the longitudinal response, and ii) two-body
currents are important in the reaction mechanism, since they were
found to have an enhancing effect of about 20$\%$ on the Euclidean
transverse response. The strong FSI effect on $R_L$ of $^4$He was
confirmed in the above mentioned LIT calculation \cite{EfL97,EfL98}
with a semirealistic force, where reliable results were obtained not
only for a transform of $R_L$, but also for $R_L$ itself. 
 
The main aim of the present work is the calculation of $R_T$ for $^4$He 
with a complete treatment of the final state interaction and 
using a current which is consistent with the potential.
Since it is the first calculation using the LIT
method we have chosen a semirealistic potential, i.e.~the
Malfliet-Tjon potential~\cite{MaT69}. The longitudinal response function of $^4$He will
be reconsidered with this potential as well. In our calculation
the final state interaction of the four-body continuum is fully taken
into account via the LIT method. Compared to the
Laplace transform, the use of a Lorentz kernel has the advantage that
a stable inversion of the integral transform is possible, thus
allowing a direct comparison of the theoretical predictions with
experimental data. 

In the following section we give a brief theoretical overview of the
LIT method and remind the reader of the definition of the inclusive
electron scattering response functions. Then in Sect.~\ref{results} we
discuss our results for the longitudinal and transverse response
functions, and in the last section we draw some conclusions and give
an outlook.

\section{Theoretical Framework}
In this section we  outline the main theoretical concepts
for the present calculation of the electromagnetic response functions
which govern the inclusive electron scattering off nuclei. First we
give a short account of the Lorentz integral transform (LIT)
which allows one to calculate the response functions with complete
inclusion of the nuclear interaction in the final states. Then, in the
second part, we  briefly review the definition of the
electromagnetic response functions and the model of the nuclear one-
and two-body currents. 

\subsection{The Lorentz integral transform}
As already mentioned, the basic observable is a response function
which 
in the laboratory system
has the general form
\beq
\label{frisp}
R_{{\cal O}}(\omega,{\boldsymbol q})=\int \!\!\!\!\!\!\!\sum _{f}| 
\left\langle \Psi_{f}| {\cal O}({\boldsymbol q})| 
\Psi _{0}\right\rangle |
^{2}\delta \left( E_{f}+\frac{{\boldsymbol q}^2}{2M}-E_{0}-\omega \right)\,, 
\eeq
where  
$M$ is the target mass, $| \Psi_{0/f} \rangle$ and 
$E_{0/f}$ denote initial and final state wave functions and energies,
respectively, while $\omega$ and ${\boldsymbol q}$ are the energy and momentum
transfers. The operator ${\cal O}$, describing the
perturbative excitation mechanism, will 
be specified  in the following subsection. The $\delta$-function 
ensures energy conservation. The response function includes a sum over
all  possible final states, which are excited by the electromagnetic
probe. This includes also the scattering states in the
continuum. Therefore, in a straightforward evaluation one 
would need to calculate besides the initial ground state the final excited
bound and continuum states of which the latter constitute the major
obstacle for a many-body nucleus if one wants to include the nuclear
interaction rigorously. 

In the LIT method~\cite{EfL94} this difficulty is
circumvented  by considering instead of the response function
$R_{\cal O}(\omega,{\boldsymbol q})$ an integral transform $L_{\cal
O}(\sigma,{\boldsymbol q})$ with a Lorentzian kernel defined for a complex
parameter $\sigma=\sigma_R+i\,\sigma_I$ by
\begin{equation}
L_{\cal O}(\sigma,{\boldsymbol q})=\int d\omega
\frac{R_{{\cal O}}(\omega,{\boldsymbol q})}
{(\omega-\sigma_R)^{2}+\sigma_I^{2}}\,.\label{lorentz_transform} 
\end{equation}
The imaginary part of $\sigma$ is kept at a constant finite value
($\sigma_I\ne 0$) determining the resolution of the
integral transform. The basic idea of considering
the integral transform is that it can be evaluated from the norm of a
state $|\widetilde{\Psi}_{\sigma,{\boldsymbol q}}^{{\cal O}}\rangle$, i.e.
\beq
\label{lit}
L_{\cal O}(\sigma,{\boldsymbol q})=
\langle\widetilde{\Psi}_{\sigma,{\boldsymbol q}}^{{\cal O}}
|\widetilde{\Psi}_{\sigma,{\boldsymbol q}}^{{\cal O}}\rangle\,,
\eeq
where $|\widetilde{\Psi}_{\sigma,{\boldsymbol q}}^{{\cal
O}}\rangle$ is the unique solution of the inhomogeneous 
``Schr\"{o}dinger-like'' equation 
\begin{equation}
(H-E_{0}-\sigma)|\widetilde{\Psi}_{\sigma,{\boldsymbol q}}^{{\cal O}}
\rangle={\cal O}({\boldsymbol q})|{\Psi_{0}}\rangle\,.\label{liteq}
\end{equation}
Here $H$ denotes the nuclear Hamiltonian.
Because of the presence of a nonvanishing imaginary part $\sigma_I$ in
Eq.~(\ref{liteq}) and the fact that its right-hand side is
localized, one has an asymptotic boundary condition like for a bound
state. Thus, one can apply bound-state techniques for its
solution. The response function itself, as a function
of the energy $\omega$ for fixed ${\boldsymbol q}$, is then obtained by inverting
the integral transform (\ref{lorentz_transform}) for which various
methods have been devised~\cite{EfL99,AnL05}. 

\subsection{The response functions of electron scattering}
In the one-photon-exchange approximation, the inclusive cross section
for electron scattering off a nucleus is given in terms of two
response functions, i.e.  
\begin{equation}
\frac{d^2 \sigma}{d\Omega d{\omega}}=\sigma_M
\left[\frac{Q^4}{{\boldsymbol q}^4} { R_L(\omega,{\boldsymbol q}
    )}+\left(\!\frac{Q^2}{2 {\boldsymbol q}^2}+\tan^2{\frac{\theta}{2}}\!
  \right) {R_T(\omega,{\boldsymbol q })}\right]
\end{equation}
where $\sigma_M$ denotes the Mott cross section, $Q^2=-q_{\mu}^2={\boldsymbol
q}^2-\omega^2$ the squared four momentum transfer with $\omega$ and
${\boldsymbol q}$ as energy and three-momentum transfers, respectively, and
$\theta$ the scattering angle. The longitudinal 
and transverse response functions,
$R_L(\omega,{\boldsymbol q})$ and $R_T(\omega,{\boldsymbol q})$, are determined
by the transition matrix elements of the Fourier transforms  of the 
charge and the transverse current density operators, $\rho({\boldsymbol q})$
and ${\boldsymbol j}_T({\boldsymbol q})$, respectively,
\beqa
R_L(\omega,{\boldsymbol q})&=& \int \!\!\!\!\!\!\!
\sum_f| \left\langle \Psi
    _{f}| {\rho}({\boldsymbol q})| \Psi
    _{0}\right\rangle | ^{2}\delta
\left(E_{f}+\frac{{\boldsymbol q}^2}{2M}-E_{0}-\omega \right)\,,
\label{rl}\\
R_T(\omega,{\boldsymbol q })&=&\int  \!\!\!\!\!\!\!
\sum_f| \left\langle \Psi _{f}|{\boldsymbol j}_T({\boldsymbol q}) 
| \Psi _{0}\right\rangle | ^{2}\delta
\left(E_{f}+\frac{{\boldsymbol q}^2}{2M}-E_{0}-\omega \right)\,.
\label{rt}
\eeqa

\subsubsection{Multipole expansion}
It is useful to decompose the charge and current densities into
Coulomb, longitudinal, and transverse electric and magnetic
multipoles~\cite{EiG70} 
\beqa
{\rho}({\boldsymbol q})&=&4\pi\sum_{J\mu}
C^J_{\mu}(q)\,Y^{J}_{\mu}(\hat{q})^*\,,\label{mult_charge}\\
{\boldsymbol{j}}\left( \boldsymbol{q}\right) & = & \sum _{J\mu }\sqrt{4\pi}
\hat{J}[ L^J_{\mu }(q)\boldsymbol{e}_{0}-\frac{1}{\sqrt{2}}(T^{el,J}_{\mu }(q)
+\mu\,{T}^{mag,J}_{\mu }(q))\boldsymbol{e}^{*}_{\mu }],\label{exp2}
\eeqa
where $\{\boldsymbol{e}_{\mu};\,\mu=0,\pm 1\}$ is a set of orthogonal
spherical unit vectors with $\boldsymbol{e}_0$ along $\boldsymbol{q}$. 
Here the Coulomb and transverse multipole operators are defined  as
\beqa
C^J_{\mu}(q)&=&\frac{1}{4\pi}\int d\hat{q}' \tilde{\rho} 
({\boldsymbol q}') Y^{J}_{\mu}(\hat{q}') \,, \label{coulomb}\\ 
L^J_{\mu }(q)&=&\frac{i}{4\pi }\int d\hat{q}^{\, \prime }
(\hat{\boldsymbol q}'\cdot {\boldsymbol{j}}(\boldsymbol{q}^{\prime }))Y^{J}_{\mu}(\hat{q}') 
\,,\label{longitudinal}\\
{T}^{el,J}_{\mu }(q)&=&\frac{i}{4\pi }\int d\hat{q}^{\, \prime }\left(
\hat{\boldsymbol{q}}^{\, \prime }\times \boldsymbol{Y}^{J}_{J\mu}(\hat{q}^{\, \prime })\right) \cdot {\boldsymbol{j}}\left(
\boldsymbol{q}^{\prime }\right)\,,\label{transverse_el}\\ 
{T}^{mag,J}_{\mu }(q)&=&\frac{1}{4\pi }\int d\hat{q}^{\, \prime
}{\boldsymbol{j}}\left( \boldsymbol{q}^{\prime }\right) \cdot \boldsymbol{Y}^{J}_{J\mu}(\hat{q}^{\, \prime })\,.\label{transverse_mag} 
\eeqa
Coulomb and longitudinal multipoles are related via current
conservation, i.e.
\beq
L^J_{\mu}(q)=\frac{i}{q}\Big[H,C^J_{\mu}(q)\Big]\,.
\label{conservation} 
\eeq 

\subsubsection{Siegert theorem and Siegert operators}
As is well known, the Siegert theorem states that in the low energy
limit the transverse electric multipoles are related to the Coulomb
multipoles. This is a very important and also very useful theorem,
since it enables one to calculate, in the low energy regime, electric
transition matrix elements from the charge density, without explicit
knowledge of the current operator. 

Even beyond the low energy regime, one can always write  each  electric
multipole as a sum of a Siegert operator plus a correction term. This
is achieved by casting the transverse electric multipole operator in
(\ref{transverse_el}) into the form 
\beqa
 {T}^{el,J}_{\mu }(q) &=&-\frac{1}{4\pi }\int d\hat{q}^{\, \prime }
\left[ \sqrt{\frac{J+1}{J}} ({\hat{\boldsymbol{q}}'\cdot {\boldsymbol{j}}
(\boldsymbol{q}')}) Y^J_{\mu }(\hat{q}')+\frac{\hat{J}}{\sqrt{J}}
\boldsymbol{Y}^{J}_{J+1\mu} (\hat{q}')\cdot {\boldsymbol{j}}
(\boldsymbol{q}^{\prime })\right]\nonumber\\
&=&{S}^J_{\mu }(q) + {K}^J_{\mu }(q)\label{mtras}\,,
\eeqa
where the first term is related to the the longitudinal
multipole operator (see Eq.~(\ref{longitudinal})) 
\beq
{S}^J_{\mu }(q)=-\frac{1}{4\pi}\sqrt{\frac{J+1}{J}}\int d\hat{q}^{\,
\prime } ({\hat{\boldsymbol{q}}'\cdot {\boldsymbol{j}}
(\boldsymbol{q}')}) Y^J_{\mu }(\hat{q}')
=i\sqrt{\frac{J+1}{J}}L^J_{\mu }(q) \label{siegert1}
\eeq
and
\begin{equation}
{K}^J_{\mu }(q)=-\frac{1}{4\pi }\frac{\hat{J}}{\sqrt{J}} \int
d\hat{q}^{\, \prime } \boldsymbol{Y}^{J}_{J+1\mu}(\hat{q}^{\, \prime
})\cdot {\boldsymbol{j}}(\boldsymbol{q}^{\prime }). \label{correction}
\end{equation}
Using the relation (\ref{conservation}) from current
conservation, the first term can be expressed by the Coulomb
operator of Eq.~(\ref{coulomb}) yielding
\beq
{S}^J_{\mu }(q) = - \frac{1}{q} \sqrt{\frac{J+1}{J}}
\Big[H,C^J_{\mu}(q)\Big]\,.\label{siegert2}
\eeq
This form of ${S}^J_{\mu }(q)$ is called the ``{\em
Siegert operator}''. As is often said, the advantage of the Siegert 
operator is that, if evaluated with the one-body charge density, it
includes already the dominant part of the MEC. The additional term
${K}^J_{\mu }(q)$ is a correction to the Siegert operator, and 
in the limit that the photon momentum goes to zero, the Siegert
operator dominates the electric multipole, since the correction
${K}^J_{\mu }(q)$ is two orders in $q$ higher. Therefore, the
approximation of the transverse electric multipoles by the Siegert 
operators is quite reliable at low photon momentum, i.e.\ for $qR\ll 1$,
where $R$ characterizes the spatial extension of the system. However,
with increasing ${q}$, one has to calculate also the contribution
of ${K}^J_{\mu }(q)$, which requires the knowledge of the explicit
form of the current operator. 

\subsubsection{The current density operators}
In general, the electromagnetic current density of a nucleus can be
decomposed in a superposition of one- and many-body operators. 
In a non-relativistic approach, as is the case here, the
electromagnetic one-body four-current is given by the free nucleon
current density operator, i.e.\ 
\beq
\rho_{(1)}({\boldsymbol x}) = \frac{e}{2}\sum_k (1+\tau^3_k)\,\delta({\boldsymbol
x}-{\boldsymbol r}_k), 
\label{chargeA}
\eeq
for the charge density, where $\tau^3_k$ is the third component of the isospin of the ``$k$-th'' particle.
The 
current density consists of a convection and a spin current 
\beq
{\boldsymbol j}_{(1)}({\boldsymbol x})={\boldsymbol j}^c_{(1)}({\boldsymbol x})+{\boldsymbol j}^s_{(1)}({\boldsymbol x})\,,
\eeq
with
\beqa
{\boldsymbol j}^c_{(1)}({\boldsymbol x})&=& \frac{e}{4m}\sum_k (1+\tau^3_k)\,
\{{\boldsymbol p}_k,\delta({\boldsymbol x}-{\boldsymbol r}_k)\}\,,\nonumber\\ 
{\boldsymbol j}^s_{(1)}({\boldsymbol x})&=& i\frac{e}{2m} \sum_k 
(\mu^s+\tau^3_k\mu^v)\,{\boldsymbol \sigma}_k \times [{\boldsymbol
p}_k,\delta({\boldsymbol x}-{\boldsymbol r}_k)]\,, 
\label{currents}
\eeqa
where  $\mu^s$, $\mu^v$, ${\boldsymbol\sigma}_k$, ${\boldsymbol p}_k$, and $m$ denote
isoscalar and isovector nucleon magnetic moments,  spin,
momentum and mass of the ``$k$-th'' particle, respectively. These expressions
describe point particles. In order to take into account the internal
nucleon structure, charge and magnetic moments have to be replaced by the 
corresponding form factors 
\beq
\frac{1}{2}(1+\tau^3_k)\rightarrow
G_{E}^s(Q^2)+\tau^3_kG_{E}^v(Q^2)\,,\quad
\mu^s+\mu^v\tau^3_k\rightarrow
G_{M}^s(Q^2)+\tau^3_kG_{M}^v(Q^2)\,. 
\eeq
For on-shell particles, these form factors depend on the squared four
momentum transfer $Q^2$ alone. In principle, this is 
no longer 
true for the off-shell situation. However, in view of the fact that little
is known about the off-shell continuation and furthermore for the
moderate energy and momentum transfers considered here, the neglect
of such effects is justified. 
 Here we use for all form factors a common
dependence on $Q^2$ of the usual dipole form~\cite{GaK71}. 

In order to satisfy the continuity equation 
\beq
\boldsymbol{ \nabla } \cdot {\boldsymbol j}_{(2)}({\boldsymbol
x})=-i[V,\rho_{(1)}({\boldsymbol x})]\,. \label{continuity_eq}
\eeq 
a momentum and/or isospin dependent two-body interaction $V$ requires
a two-body current density operator, the interaction (or meson
exchange) current. As is well known, relation~(\ref{continuity_eq}) is
not sufficient to determine 
the two-body current uniquely. Therefore one needs in principle a
dynamic model for the nuclear potential which reveals the underlying
interaction mechanism. Such a model is supplied by the meson exchange
picture of the $NN$-interaction. But even for a phenomenological
potential often a consistent MEC can be constructed if the potential
has the formal appearance of a meson exchange potential. This is
achieved using the method developed independently in~\cite{Ris85} and
\cite{BuL85}. The method is based on the observation that if a
potential is given as a sum of Yukawa-like terms, one can interpret
the potential as produced by the exchange of fictitious mesons whose
exchange currents are known. In the present case we use the semirealistic
Malfliet-Tjon potential~\cite{MaT69}, which is isospin dependent and thus
requires a MEC. Despite its phenomenological character, it fulfills
the just mentioned requirement and thus it is possible to derive a
consistent MEC which satisfies (\ref{continuity_eq}) with this
potential. 

Explicitly, the MT-potential has the form
\beq
V(r)=V_{13}(r){P}_{13}+V_{31}(r) {P}_{31}\,,\label{MT_potential}
\eeq
where ${P}_{13}$ and ${P}_{31}$ denote the  projectors on the
spinsinglet-isotriplet and spintriplet-isosinglet two-nucleon states,
respectively, 
\beqa
{P}_{13}&=&\frac{1}{16}(1-{\boldsymbol \sigma}_1 \cdot {\boldsymbol
\sigma}_2)(3+{\boldsymbol \tau}_1 \cdot {\boldsymbol\tau}_2)\,, \\  
{P}_{31}&=&\frac{1}{16}(3+{\boldsymbol \sigma}_1 \cdot {\boldsymbol
\sigma}_2)(1-{\boldsymbol \tau}_1 \cdot {\boldsymbol \tau}_2)\,.  
\eeqa
Furthermore, the radial functions are given by
\beqa
V_{13}(r)&=& 4\pi(A J_{m_1}(r)-B J_{m_2}(r)), \\
V_{31}(r)&=& 4\pi(C J_{m_1}(r)-D J_{m_2}(r)), 
\eeqa
with $J_{m_{1/2}}(r)=e^{-m_{1/2}r}/4\pi r$,
$m_1=3.11$~fm$^{-1}$, $m_2=1.55$~fm$^{-1}$
and $A=C=1458.047$ MeV~fm, $B=520.872$ MeV~fm, $D=635.306$ MeV~fm.
Exhibiting explicitly the spin-isospin dependence, one can rewrite the
potential in the form
\beq
V(r)=V_0(r)+V_\sigma(r)\,{\boldsymbol \sigma}_1 \cdot {\boldsymbol
\sigma}_2+(V_\tau(r)+V_{\sigma\tau}(r)\,{\boldsymbol \sigma}_1 
\cdot {\boldsymbol \sigma}_2){\boldsymbol \tau}_1 
\cdot {\boldsymbol \tau}_2\,,\label{potential}
\eeq
with
\beqa
V_0(r)&=&\frac{12 \pi}{16} \Big(AJ_{m_1}(r) -BJ_{m_2}(r)
+CJ_{m_1}(r)-DJ_{m_2}(r)\Big)\,,\\ 
V_\sigma(r)&=&\frac{4 \pi}{16} \Big(-3AJ_{m_1}(r) +3BJ_{m_2}(r)
+CJ_{m_1}(r)-DJ_{m_2}(r)\Big)\,,\\ 
V_\tau(r)&=&\frac{4 \pi}{16} \Big(AJ_{m_1}(r) -BJ_{m_2}(r)
-3CJ_{m_1}(r)+3DJ_{m_2}(r)\Big)\,,\\ 
V_{\sigma\tau}(r)&=&\frac{4 \pi}{16}\Big(-AJ_{m_1}(r)+BJ_{m_2}(r)
-CJ_{m_1}(r)+DJ_{m_2}(r)\Big)\,.
\eeqa
Because the isospin independent part commutes with the charge
operator, only the isospin dependent part of (\ref{potential}) is
relevant for the MEC. One can bring it into the form
\beq
V_\tau(r)+V_{\sigma\tau}(r)=4\pi\, {\boldsymbol \tau}_1 \cdot
 {\boldsymbol \tau}_2 \Big( \alpha J_{m_1}(r) + \beta
 J_{m_2}(r)+( \gamma J_{m_1}(r) + \delta J_{m_2}(r)) 
{\boldsymbol \sigma}_1 \cdot {\boldsymbol \sigma}_2 \Big)\,,
\eeq
where the new constants are defined as
\begin{eqnarray}
\nonumber
&&\alpha=\frac{1}{16}(A-3C), \qquad
\beta=-\frac{1}{16}(B-3D),\\
\nonumber
&&\gamma=-\frac{1}{16}(A+C), \qquad
\delta=\frac{1}{16}(B+D).
\end{eqnarray}
Now, since $J_{m}(r)$ represents the propagator of an exchanged scalar
meson of mass $m$, the corresponding meson exchange current is easily
constructed (for details see Ref.~\cite{Bac05}) and one finds
\beqa
{\boldsymbol j}^{\rm MT,I-III}_{(2)} ({\boldsymbol x},{\boldsymbol r}_1,{\boldsymbol r}_2)&=&
4\pi ({\boldsymbol \tau}_1 \times{\boldsymbol \tau}_2)_3\Big[\alpha
J_{m_1}(|{\boldsymbol r}_1- {\boldsymbol x}|) \stackrel{\leftrightarrow}
{\boldsymbol \nabla}_xJ_{m_1}(|{\boldsymbol x}- {\boldsymbol r}_2|) +\beta
J_{m_2}(|{\boldsymbol r}_1-{\boldsymbol x}|) \stackrel{\leftrightarrow}{\boldsymbol
\nabla}_x J_{m_2}(|{\boldsymbol x}- {\boldsymbol r}_2|)\nonumber\\ 
&&+\Big(\gamma J_{m_1}(|{\boldsymbol r}_1- {\boldsymbol x}|) \stackrel{\leftrightarrow}{\boldsymbol \nabla}_x J_{m_1}(|{\boldsymbol x}-{\boldsymbol r}_2|)+\delta J_{m_2}(|{\boldsymbol r}_1- 
{\boldsymbol x}|) \stackrel{\leftrightarrow}{\boldsymbol\nabla}_x J_{m_2}
(|{\boldsymbol x}-{\boldsymbol r}_2|)\Big) ~{\boldsymbol \sigma}_1 \cdot {\boldsymbol
\sigma}_2 \Big].\label{mecx}
\eeqa
It is a simple task to prove that this current fulfills the continuity
equation (\ref{continuity_eq})  with the Malfliet-Tjon potential in
(\ref{MT_potential}). For later purpose we list also the Fourier
transform of the above MEC 
\begin{eqnarray}
{\boldsymbol j}^{\rm MTI-III}_{(2)} ({\boldsymbol q}, {\boldsymbol r}_1, {\boldsymbol r}_2)&=&
\frac{e^{i{\boldsymbol q}\cdot {\boldsymbol R}}}{\pi^2} 
({\boldsymbol \tau}_1 \times{\boldsymbol \tau}_2)_3
\Big[\alpha {\boldsymbol \nabla}_{{\boldsymbol r}} I_{m_1}({\boldsymbol q},{\boldsymbol r})
+\nonumber\\
&&\beta {\boldsymbol \nabla}_{{\boldsymbol r}} I_{m_2}({\boldsymbol q},{\boldsymbol r})
+{\boldsymbol \sigma}_1 \cdot {\boldsymbol \sigma}_2
\Big(\gamma{\boldsymbol \nabla}_{{\boldsymbol r}} I_{m_1}({\boldsymbol q},{\boldsymbol r})+
\delta {\boldsymbol \nabla}_{{\boldsymbol r}} I_{m_2}({\boldsymbol q},{\boldsymbol r})
\Big)\Big],
\label{2b_curr_q}
\end{eqnarray}
where ${\boldsymbol r}$ and  ${\boldsymbol R}$ denote relative and center-of-mass
coordinates of the two-body sub-system, respectively, defined as
\begin{equation}
\nonumber
{\boldsymbol r}= {\boldsymbol r}_1- {\boldsymbol r}_2\quad\mbox{and}\quad
{\boldsymbol R}= \frac{1}{2}({\boldsymbol r}_1+ {\boldsymbol r}_2)\,,
\end{equation}
and
\beq
\label{I_func}
I_{m_{1/2}}({\boldsymbol q},{\boldsymbol r})=\int d^3p\frac{e^{i{\boldsymbol r}\cdot{\boldsymbol p}}}
{[({\boldsymbol p}+\frac{1}{2}{\boldsymbol q})^2+m^2_{1/2}]
[({\boldsymbol p}-\frac{1}{2}{\boldsymbol q})^2+m^2_{1/2}]}\,.
\eeq
Here, we would like to recall that in contrast to   
a pseudoscalar meson exchange current the current in 
Eq.~(\ref{2b_curr_q})
does not include any contact term, but only  meson in flight
contributions, due to the fact that the MTI-III potential represents 
the exchange of scalar mesons. 
The functions $I_{m_{1/2}}$ contain, in fact, the  propagators
of scalar mesons with effective masses $m_1$ and
$m_2$ (for details see Ref.~\cite{FaA76}).

\section{Results and discussion}\label{results}

Now, we  present results for the longitudinal and transverse
electric and magnetic response functions for $^4$He. We have solved
the LIT-equation~(\ref{liteq}) by using an expansion of the internal
wave functions in terms of hyperspherical harmonics (HH). For the
antisymmetrization we have used the powerful algorithm developed
by Barnea and Novoselsky~\cite{BaN97}. The HH-expansion has been truncated
beyond a maximum value $K_{max}$ of the HH grand-angular
momentum. Furthermore, the convergence of the HH expansion has been
improved considerably by introducing the above mentioned effective
interaction with hyperspherical harmonics
(EIHH-approach)~\cite{BaL00}, where the original bare potential has been
replaced by an effective potential constructed via the Lee-Suzuki
method~\cite{SuL80}. When convergence is reached, the results agree
with the ones obtained with the bare potential~\cite{BaL00}. As
original potential we have used the Malfliet-Tjon MTI-III
potential~\cite{MaT69}. We found that for the ground
state good convergence is reached already with $K_{max}=10$, yielding
a binding energy of $E_0=-30.57$ MeV (the small difference with
respect to the previously published value in \cite{BaL00} is due to
the different $K_{max}$ used). For the state
$\widetilde{\Psi}_{\sigma,{\boldsymbol q}}^{{\cal O}}$ of Eq.~(\ref{lit})
we have used $K_{max}=11$ or $12$ depending on the odd or even parity of the
excitation operator $\cal{O}$, respectively. The results we are
presenting are fully converged in $K_{max}$. We begin the discussion
with the longitudinal response function.

\subsection{The Longitudinal Response Function}

\begin{figure}[htb!]
\centering
\includegraphics[scale=.5]{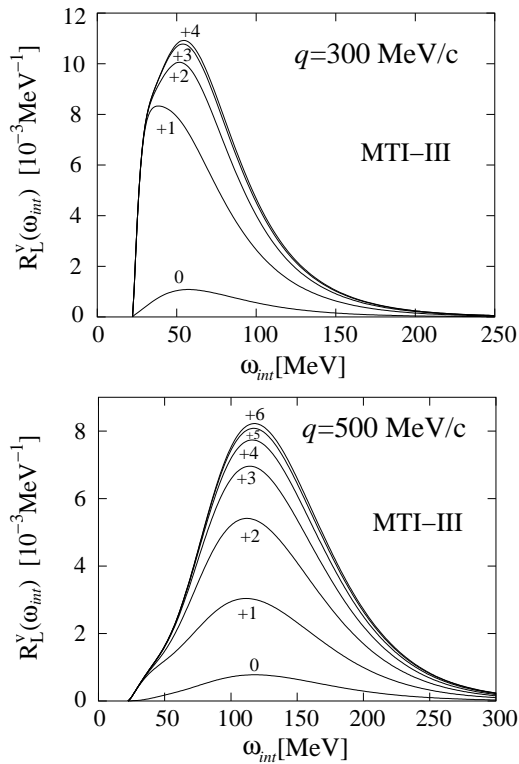}\\
\caption{Response functions of the lowest isovector Coulomb multipoles,
starting with the monopole and consecutively adding higher multipoles
up to $J_{max}=4$ for ${q}=300$~MeV/c (upper panel) and $J_{max}=6$ for
${q}=300$~MeV/c (lower panel), as a function of the intrinsic energy.} 
\label{figure1}
\end{figure}

Since the longitudinal response function is an incoherent sum of the
separate multipole contributions it is useful to expand the charge
operator into Coulomb multipoles as defined in~(\ref{mult_charge}). 
For each multipole operator the LIT-equation with ${\cal O}({\boldsymbol
q})=C^J_{\mu}(q)$ is solved. In view of the fact that isospin is a good
quantum number and that the ground state of $^4$He has total isospin
zero, one can treat the isoscalar and isovector parts of the charge
operator separately. 
The multipole expansion has been truncated at a maximum value $J_{max}$
determined by the requirement that convergence is
achieved. It is known that with increasing momentum transfer $J_{max}$
has to be increased too in order to reach convergence.

In Fig.~\ref{figure1}, we present the response functions of the isovector
multipoles, obtained from the inversion of the LIT of each multipole, as a function of the intrinsic energy $\omega_{int}=E_f -E_0$. 
We show results for two momentum
transfers $q=300$ and 500~MeV/c. One readily notes that for the lower
momentum transfer five   multipoles are sufficient in order to reach convergence,
while for the higher momentum transfer two additional multipoles need to be
included. The dominant feature is the quasi-elastic peak located at
$\omega_{int}\simeq q^2/2m$, where $m$ is the mass of the nucleon, with a shoulder on the low energy side. The latter
decreases rapidly with increasing momentum transfer. 
\begin{figure}[htb!]
\centering
\includegraphics[scale=.5]{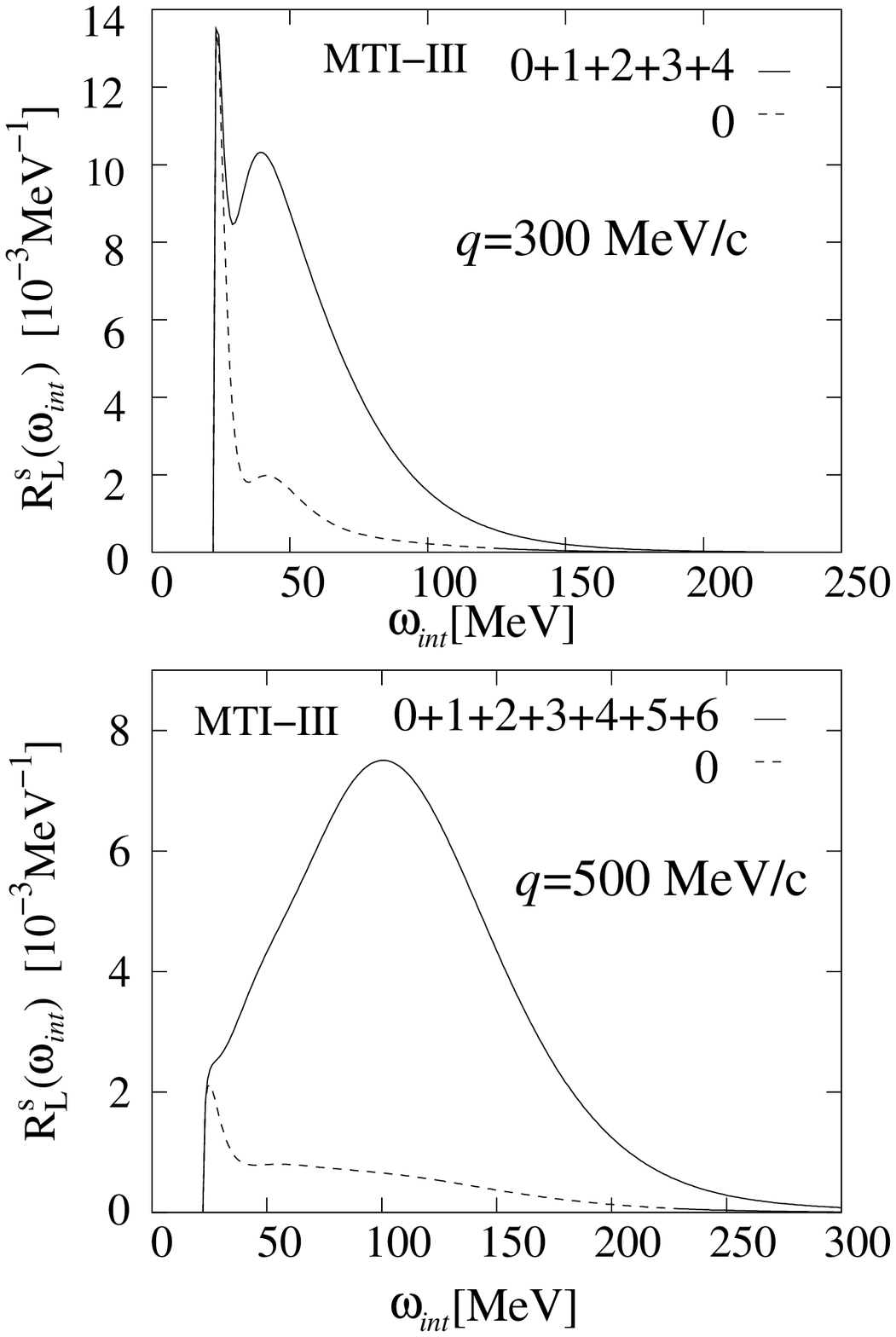}\\
\caption{The response functions of the isoscalar Coulomb monopole and
the summed multipoles up to $J_{max}=4$ for ${q}=300$~MeV/c (upper
panel) and $J_{max}=6$ for ${q}=300$~MeV/c (lower panel) as a function
of the intrinsic energy.}
\label{inv_RL}
\end{figure}

The corresponding isoscalar response functions are displayed in
Fig.~\ref{inv_RL}. The contributions of the isoscalar monopole and of
the sum of all multipoles up to $J_{max}$ are shown separately. An
interesting feature of the monopole is the pronounced peak close to
threshold which is even higher than the quasi-elastic peak for the
lower momentum transfer, but suppressed for the higher momentum
transfer, although still visible as a shoulder. In
order to better determine the width of the monopole excitation of $^4$He  
a more detailed study of low-$q$ transitions with a
realistic nuclear interaction is necessary. 

\begin{figure}[htb!]
\centering
\includegraphics[scale=.5]{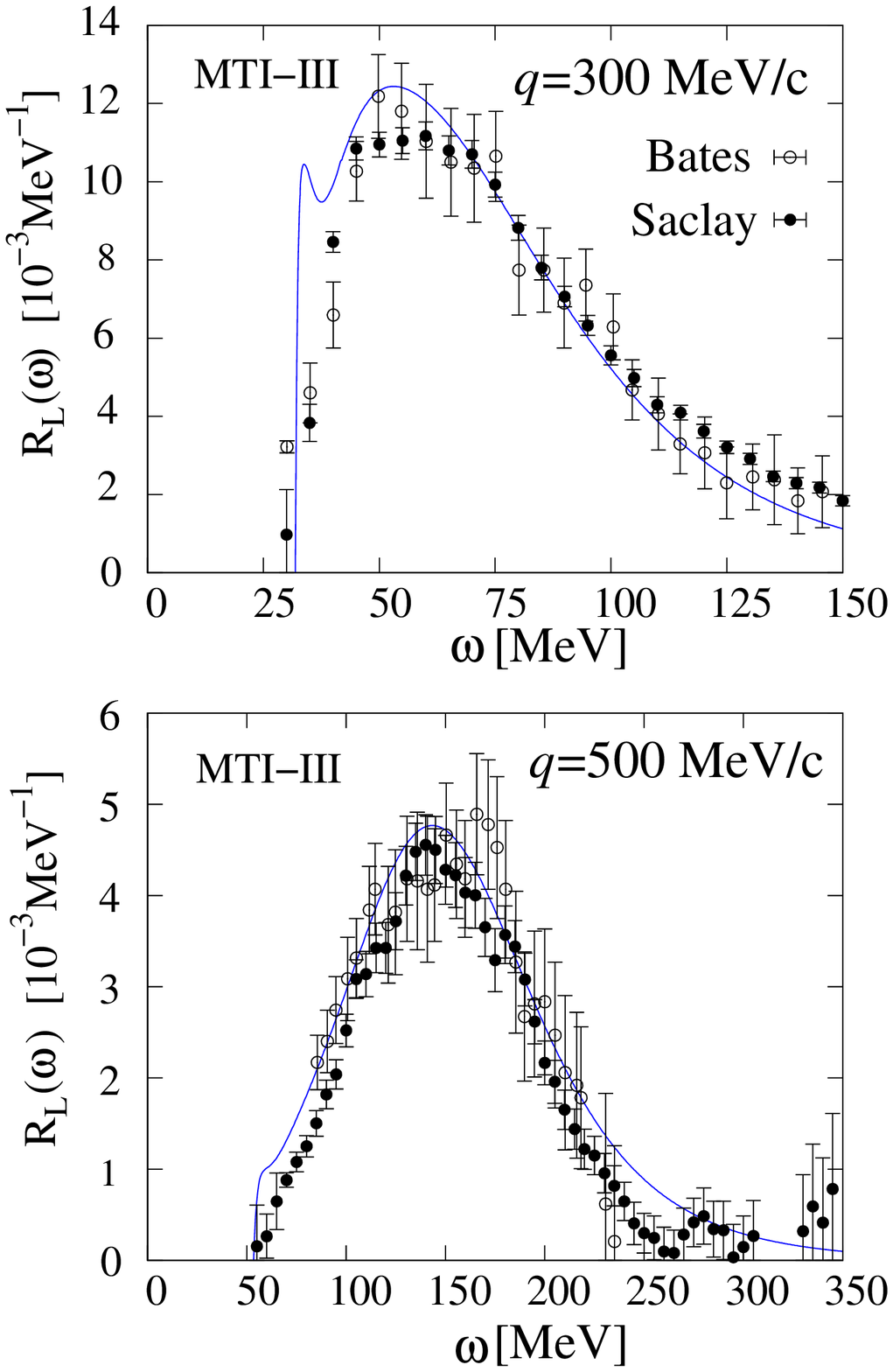}\\
\caption{(Color online) Longitudinal response function with the
MTI-III potential as function of the  energy for momentum
transfers $q=300$~MeV/c (upper panel) and  $500$~MeV/c (lower
panel). Experimental data from MIT-Bates~\cite{Dyt88} (open circles)
and Saclay~\cite{Zgh94} (filled circles).} 
\label{fig3}
\end{figure}

After the inclusion of the nucleon form factors and summing up the
isoscalar and isovector contributions, we obtain the total
longitudinal response function for varying energy transfers in the
laboratory frame, where the recoil energy of the nucleus has been
included. In Fig.~\ref{fig3}, we compare our results with the
available experimental data for momentum transfers $q=300$ and
$500$~MeV/c. We have taken into account the
Darwin-Foldy lowest order relativistic correction of the one-body
charge density by a proper modification of the nucleon charge form
factor \cite{friar}. While the correction is negligible for $q=300$
MeV/c, it leads to a damping  of about $6\%$ of the total longitudinal
strength in the quasi-elastic peak for $q=500$ MeV/c. The less
important spin-orbit relativistic correction has been neglected (see
e.g. \cite{EfL04}). 

As in the
previous calculation of $R_L$ using the Trento (TN)
potential~\cite{EfL97}, one notes that a semirealistic interaction, in
this case the MTI-III, leads to quite a good overall description of the
response in comparison to the experimental data from
Bates~\cite{Dyt88} and Saclay~\cite{Zgh94}. 

The only difference to the
previous calculation  with the TN potential is the pronounced 
peak close to threshold in case of $q=300$~MeV/c which originates from
the monopole excitation of $^4$He. On the other hand, such a peak is
not seen in the data. But it is not clear whether the experimental
energy resolution was sufficient to resolve such a structure.
It is worthwhile to mention
that a $0^+$ resonance at 20.10$\pm$0.05 MeV with a width of 270$\pm$50 keV
was determined in an electron scattering experiment at
momentum transfers $q<100$ MeV/c \cite{Wal70}. Here we do not
calculate these low-$q$ kinematics, the resonance is
very close to the "quasi-elastic peak" and quite small in size in comparison.
A much more detailed study than the present calculation would be necessary to
resolve such a rather complicated low-energy structure.

\subsection{The Transverse Response Function}

As done for the charge operator, we have expanded
the transverse current operator into electric and magnetic multipoles
according to Eqs.~(\ref{transverse_el}) and (\ref{transverse_mag}),
separating them further into isoscalar and isovector parts, because the
response function is an incoherent sum of these various multipole
contributions. 
As discussed above, the transverse current includes one- and two-body
operators. We first consider the one-body current alone, i.e.\ the spin
and the convection current of Eq.~(\ref{currents}). Later we will
add the consistent two-body current.

\subsubsection{One-body current}

\begin{figure}[htb!]
\centering
\includegraphics[scale=.5]{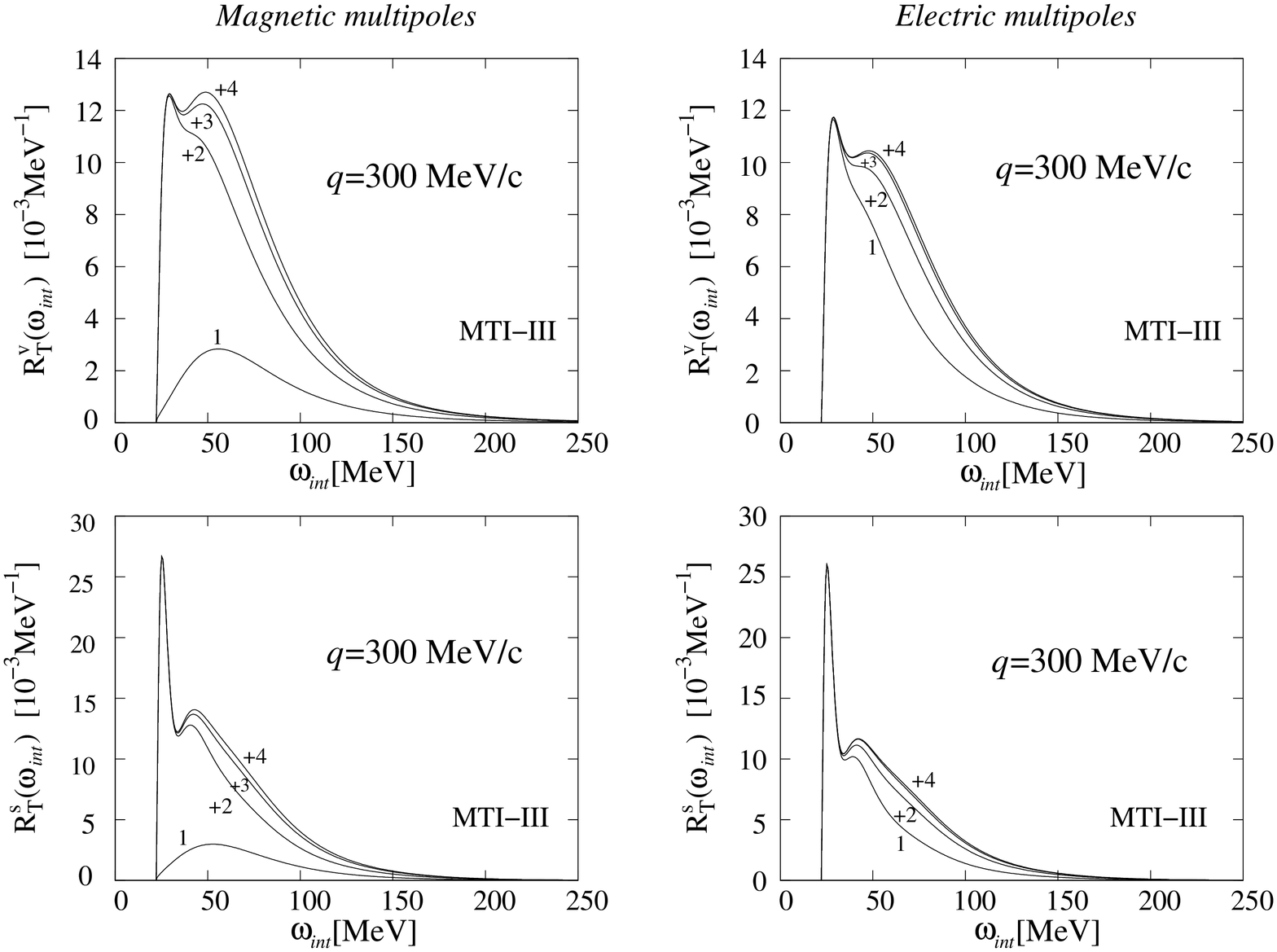}\\
\caption{Transverse response functions at
$q=300$~MeV/c as functions of the intrinsic energy of various
isovector (upper panels) and isoscalar (lower panels) magnetic (left)
and electric (right) multipoles of the spin current, starting with the
dipole and consecutively adding higher multipoles up to $J_{max}=4$.} 
\label{fig4}
\end{figure}

It is known from standard PWIA calculations, that the spin current
dominates the transverse response function at medium momentum transfers
in the region of the quasi-elastic peak. Therefore, we start the
discussion of the transverse response function of the spin current
alone. In Fig.~\ref{fig4}, we present the isoscalar and
isovector response functions of the magnetic and
  electric multipoles up to
$J_{max}=4$ for the spin current at a momentum transfer
$q=300$~MeV/c. One readily notes that the dominant  isovector
contribution exhibits a substructure in the quasi-elastic peak region,
which originates mainly from the magnetic quadrupole and
electric dipole contributions. 
A very pronounced peak is also found in the magnetic
  and electric isoscalar responses close to
threshold, which is, however, suppressed in relation to the isovector
part of the total response function after the inclusion of the nucleon
form factors, because of the smallness of the isoscalar magnetic form
factor relatively to the isovector one.

\begin{figure}[htb!]
\centering
\includegraphics[scale=.6]{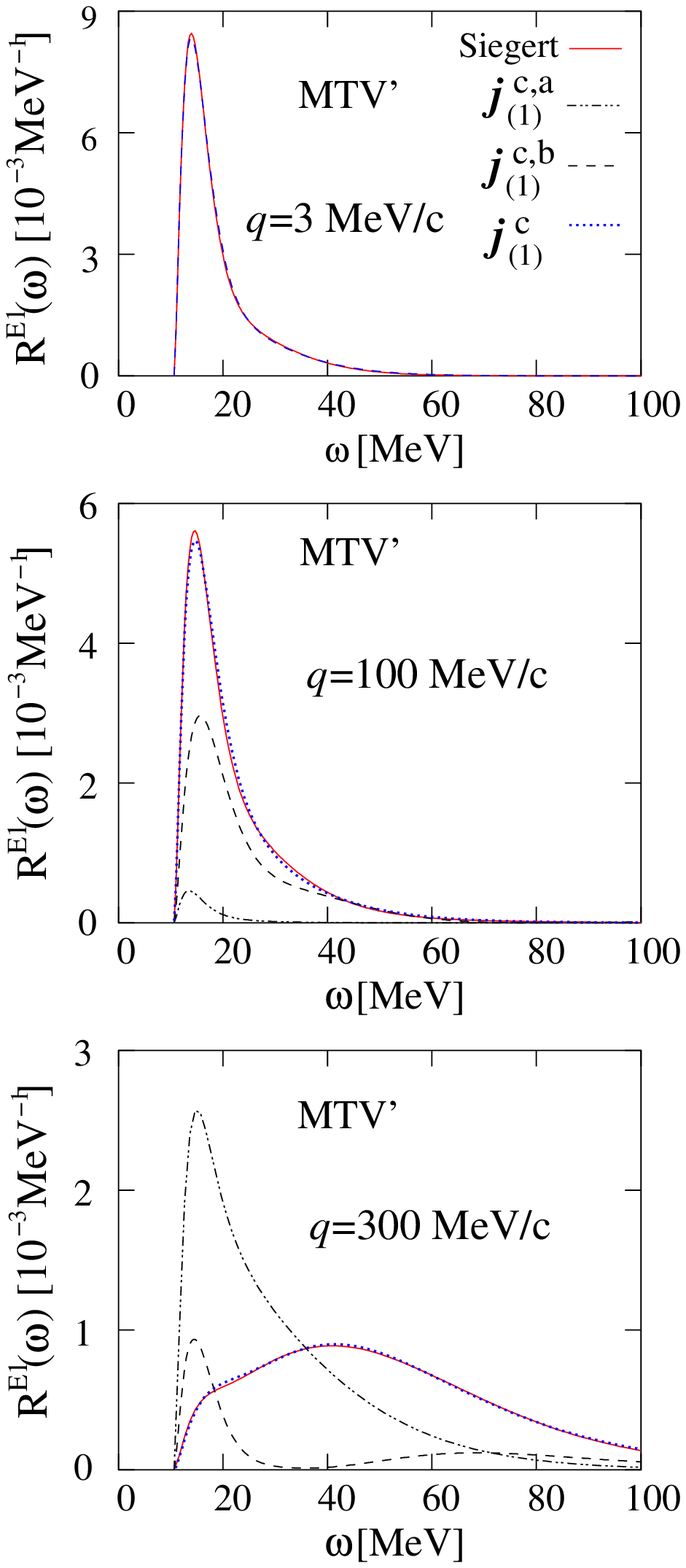}
\caption{(Color online) Siegert part of the dipole response function
for different values of the momentum transfer and evaluated with (i)
the Siegert operator of Eq.~(\ref{siegert2}) and (ii)
the operator of Eq.~(\ref{siegert1}) with the convection currents
${\boldsymbol j}^{c,a}_{(1)}({\boldsymbol q})$ and ${\boldsymbol j}^{c,b}_{(1)}({\boldsymbol q})$,
and their sum (see text) for the MTV' potential.} 
\label{fig5}
\end{figure}
The second contribution of the one-body current to the transverse
response function is given by the convection current of
Eq.~(\ref{currents}). In this case one has a derivative term caused by
the dependence on the nucleon momentum ${\boldsymbol p}_k$ in the
anticommutator, which acts on the wave function. In the EIHH approach
this derivative can be calculated analytically, due to the fact
that the radial wave function is given in terms of Laguerre and
Jacobi polynomials with hyperradial and hyperspherical variables,
respectively (see Ref.~\cite{BaL00}). In order to prove the correct
implementation of the derivative term, we make use of the Siegert
operator as a check. Firstly, we write the convection current in
momentum space as 
\begin{equation}
\label{conv}
{\boldsymbol j}^c_{(1)}({\boldsymbol q})={\boldsymbol j}^{c,a}_{(1)}({\boldsymbol q})+{\boldsymbol
  j}^{c,b}_{(1)}({\boldsymbol q})\, ,
\end{equation}
where 
\begin{eqnarray}
\label{conva}
{\boldsymbol j}^{c,a}_{(1)}({\boldsymbol q})&=&
\frac{e{\boldsymbol q}}{4m}\sum_k (1+\tau^3_k)\,e^{i{\boldsymbol q}\cdot {\boldsymbol r}_k}\,,\\
\label{convb}
{\boldsymbol  j}^{c,b}_{(1)}({\boldsymbol q})&=&
 \frac{e}{2m}\sum_k (1+\tau^3_k)\,e^{i{\boldsymbol q}\cdot {\boldsymbol r}_k} {\boldsymbol p}_k\,.
\end{eqnarray}
In this way, the first part ${\boldsymbol j}^{c,a}_{(1)}({\boldsymbol q})$ does not
contain any derivative term and is purely longitudinal, while the
second part ${\boldsymbol j}^{c,b}_{(1)}({\boldsymbol q})$ contains a longitudinal and
a transverse, term depending on the direction of  ${\boldsymbol p}_k$ with
respect to  ${\boldsymbol q}$. The check consists in the
comparison of the Siegert operator of (\ref{siegert2}) with the
operator in (\ref{siegert1}) using the convection current of
(\ref{conv}). For the case of 
a purely central potential, where no meson exchange currents are
present, the two contributions have to coincide.  As such an
interaction we take for a test
the model MTV', a modified version of the central Malfliet-Tjon-V,
as defined in \cite{Bac05}.

The results of this check for the dipole response
are exhibited in Fig.~\ref{fig5} for different momentum transfers.
First of all one can note that the dipole strength is differently
distributed depending on the momentum transfer: for the two lower
$q$-values the response function has a pronounced and quite small peak
located a few MeV above threshold, while for the highest momentum transfer,
$q=300$~MeV/c, the peak becomes considerably broader, but smaller in
height and shifted to higher energies. This different behavior has its
origin in a different dependence of the two
contributions, ${\boldsymbol j}^{c,a}_{(1)}({\boldsymbol q})$ and ${\boldsymbol
j}^{c,b}_{(1)}({\boldsymbol q})$, to (\ref{siegert1}) on the momentum
transfer. The current ${\boldsymbol j}^{c,b}_{(1)}({\boldsymbol q})$ is
dominant at $q=3$~MeV/c, while for the two higher $q$-values also the
second component, ${\boldsymbol j}^{c,a}_{(1)}({\boldsymbol q})$, becomes sizeable and
interferes constructively with ${\boldsymbol j}^{c,a}_{(1)}({\boldsymbol q})$ at
$q=100$~MeV/c but destructively at $q=300$~MeV/c. However, their total
contribution agrees perfectly well with the Siegert dipole response.  

\begin{figure}[htb!]
\centering
\includegraphics[scale=.4]{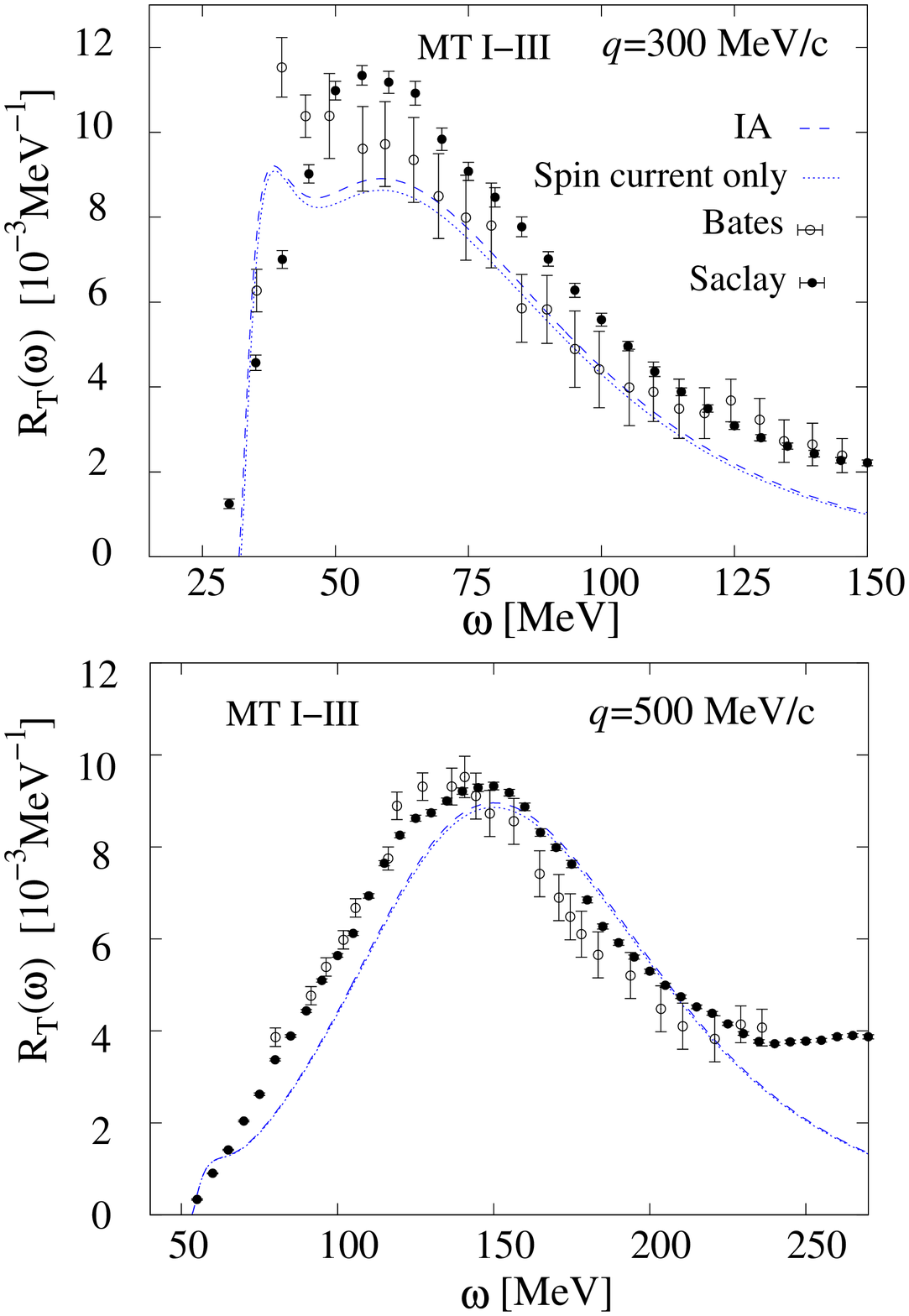}\\
\caption{
(Color online) Transverse response function in impulse
approximation  (IA) and response of the one-body spin current alone as a
function of the energy for momentum transfer ${q}=300$
(upper panel) and ${q}=500$~MeV/c (lower panel) in comparison with 
data from Bates~\cite{Dyt88} and Saclay~\cite{Zgh94}.} 
\label{fig6}
\end{figure}

Again, the convection current is expanded into electric and magnetic
multipoles,  and then the response function is calculated  for each
multipole until convergence is reached. For example, in case of medium
momentum transfers like $q=300$~MeV/c we obtain a
rather fast convergence pattern with $J_{max}=4$, since already only
the dipole and quadrupole contributions 
cover about $80\%$ of the total strength of the convection current. We
also observe that only the electric multipoles are relevant, while
the magnetic multipoles are negligible. For example, in case of
$q=300$~MeV/c, the magnetic dipole induced by the convection current
is 10 times smaller than the electric dipole. 

Summing up the total contributions of spin and convection currents, we
obtain the transverse response function in impulse approximation. In
Fig.~\ref{fig6}, we present the results for the MTI-III potential in
comparison to the experimental data from Bates \cite{Dyt88}  and
Saclay \cite{Zgh94} for momentum transfers $q=300$ and
500~MeV/c. We also show the response given by the spin current only,
which clearly dominates for both values of momentum transfer. For
$q=300$ MeV/c the addition  of the convection current produces an
enhancement of  $3-4 \%$ near the peak energy $\omega=60$
MeV. In case of $q=500$~MeV/c, the convection current contribution  is
negligible. Qualitatively, the IA results reproduce the
position and the width of the quasi-elastic peak, but show some
deviations in detail. 
For example, for $q=300$~MeV/c in the region of the maximum and above 
the theory is systematically below the data: in the maximum it is about
$10\%$ lower compared to the Bates data and $20\%$ with respect to the
Saclay data.  On the low energy side the theory exhibits a
steeper rise than the data and in addition a secondary maximum which
is not seen in the Saclay data, but might possibly be indicated by
the Bates data. 
We would like to point out that evidence for an M2 resonance at 24 MeV
was found in the above mentioned $(e,e')$-experiment \cite{Wal70}.
On the other hand, for $q=500$~MeV/c, the
theoretical height of the quasi-elastic peak is only about $5\%$ lower
with respect to experiment, but the peak position is shifted by about 
$6$~MeV towards higher energies. Furthermore, the theoretical shoulder
near the threshold is barely seen in the data. These substructures at
low energies arise from the electric dipole and magnetic quadrupole
contributions of the spin current. 
 
It is conceivable that the overall
missing strength could be provided by the up to now neglected two-body
currents. Another explanation could be related to some inadequacies of
the semirealistic description of the nuclear Hamiltonian
or to relativistic effects. 
In the following subsection we will investigate the first possibility. 

\subsubsection{Two-body current}

\begin{figure}[htb!]
\centering
\includegraphics[scale=.5]{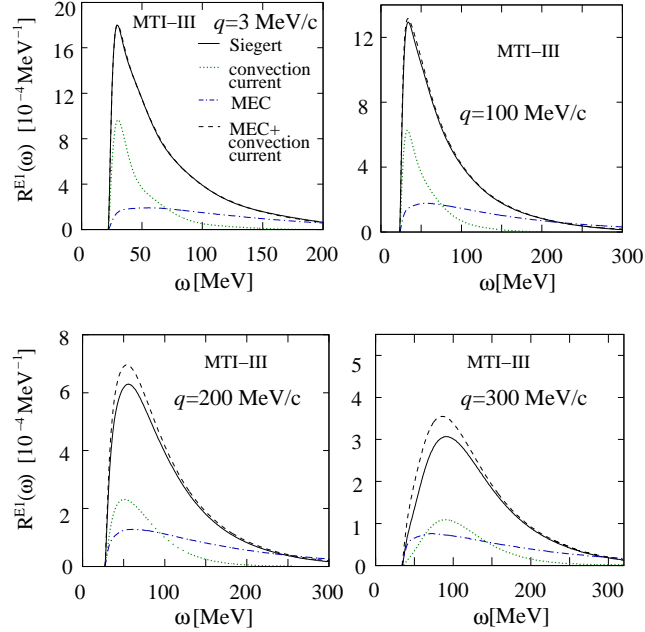}
\caption{(Color online) Dipole response function with the MTI-III
potential, as a function of the  energy and for different
values of the momentum transfer. The response is evaluated for (i)
the Siegert operator of (\ref{siegert2}) (solid curves) and for the
operator of (\ref{siegert1}) with (ii) convection current (dotted),
(iii) MEC (dash-dotted) and (iii) their sum (dashed).}
\label{fig7}
\end{figure}

The consistent two-body current for the MTI-III potential,
${\boldsymbol j}^{\rm MTI-III}_{(2)} ({\boldsymbol q}, {\boldsymbol r}_1, {\boldsymbol r}_2)$ is given
in (\ref{2b_curr_q}). Its general multipole decomposition is listed in
Eq.~(\ref{mec_multipole}) of the appendix. 
In the present 
MEC calculation we neglect the center of mass motion
of the two-body subsystem, i.e.\ we set $e^{i{\boldsymbol q}\cdot
{\boldsymbol R}}=1$.
We would like to point out that in this approximation all magnetic and
even electric multipoles vanish identically (see appendix). One
expects this approximation to be quite good for small momentum
transfers, i.e.\ $qR \leqq 1$. In Fig.~\ref{fig7}, we compare, for the
most important dipole contribution, results with explicit MEC and
convection current on the one hand and with the Siegert operator 
on the other hand.
In contrast to the MTV case depicted in Fig.~\ref{fig5} with the
MTI-III potential, the Siegert dipole does not agree 
with the response of the Siegert part of the convection current alone,
due to the presence of the MEC. At low momentum transfer the
convection current leads to a response, whose peak height is 
only about half of the Siegert peak, while at higher
momentum transfer the effect of MEC becomes even stronger. The dipole
response function of the MEC is also shown in
Fig.~{\ref{fig7}}. It is interesting to note that the
strength of the convection current is located near the quasi-elastic
region whereas the MEC has a much broader distribution.
Taking as excitation operator the sum of convection current and MEC,
we get a perfect agreement with the Siegert dipole for $q=3$~MeV/c,
which clearly proves that the approximation introduced in the two-body current
is reliable at low momentum transfer, as expected. For $q=100$~MeV/c
our approximation can still be considered very good.
However at the higher values of momentum transfer, one notes
an increasing difference between the two evaluations which has to be
assigned to the neglected c.m.\ motion of the two-body subsystems in
the calculation of the MEC. One further observes that the effect of
the approximation consists mainly in an enhancement the peak height of
the response, while position and form of the peak are not affected.
The peak reduction amounts to about $10 \%$  and $20 \%$ for
$q=200$ and 300~MeV/c, respectively. 

\begin{figure}[htb!]
\centering
\includegraphics[scale=.5]{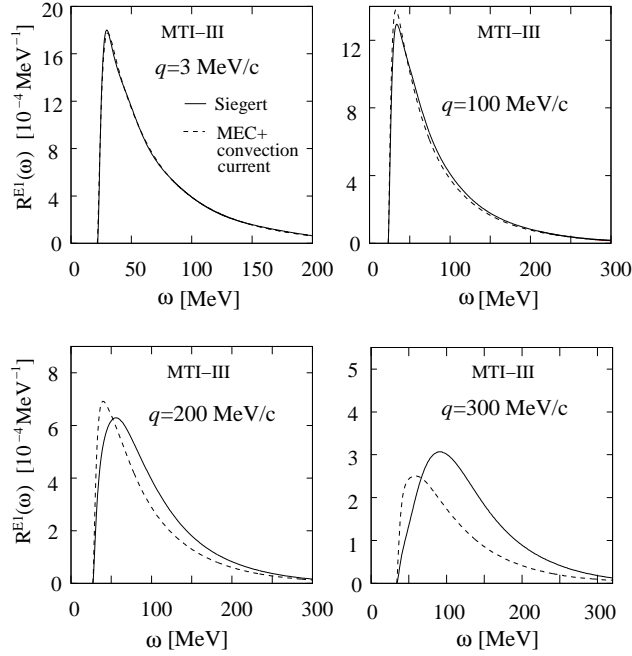}
\caption{Dipole response  function with the MTI-III potential as
  a function of the  energy, for different values
  of the momentum transfer.
The response is 
 evaluated   for  (i) the
Siegert operator of (\ref{siegert2}) (solid) and (ii) 
for total dipole 
operator according to (\ref{mtras}) with the sum of  convection current
with MEC (dashed).}
\label{fig8}
\end{figure}
Furthermore, we can check to what extent the Siegert operator alone is
a good estimate of the total dipole operator, i.e.\ whether the
correction term of (\ref{correction}) can safely be neglected.
To this purpose we consider the total multipole of Eq.~(\ref{mtras}),
where  the correction term (\ref{correction}) has been included in
addition to the Siegert part (\ref{siegert1}), and we compare it to
the Siegert operator itself alone. In Fig.~\ref{fig8}, we show the
result of this check again for the isovector dipole response as a
function of the energy for different momentum
transfer values. The total dipole response function induced by the
convection current and by the MEC with mutual interference leads to
the same result as the Siegert dipole response for the very low momentum
transfer $q=3$ MeV/c, and even up to a momentum transfer of
$q=100$~MeV, the Siegert operator is still a reliable approximation. This is
not anymore the case at higher momentum transfers where one readily
notes an increasing difference between the total dipole response and
the Siegert one alone. For momentum transfers $q=200$ and
300~MeV/c, the Siegert operator clearly underestimates
the total dipole response at energies below the peak and
overestimates it in the  tail. In fact, the strength of the Siegert
response is moved as a whole towards higher energies with a shift of
the peak position with respect to the total dipole response by about
$17$ MeV and 33~MeV for $q=200$ and 300~MeV/c, respectively. On the
other hand, with respect to the total dipole response the peak height
of the Siegert response is reduced by about $10\%$ at $q=200$~MeV/c
and enhanced by about $20\%$ for $q=300$~MeV/c. Obviously, the Siegert
operator is not a good approximation for the total dipole response for
$q> 100$ MeV/c. At $\omega=100$~MeV, for example, compared to
the result of the total dipole response, the Siegert response 
leads to an overestimation of $35 \%$ and $70\%$ for $q=200$ and
300~MeV/c, respectively. 

Though the situation may be quite different for other electric
multipoles, like quadrupole and octupole, and even more for the
isoscalar multipoles, we think that the results of this check on the
dominant isovector dipole clearly suggests that one needs to consider
the total electric multipoles when investigating the transverse
response function at medium to high momentum transfers, since in this
kinematic region the Siegert approximation is not reliable
anymore. Thus, an explicit knowledge of the two-body current operator
is necessary. 

\begin{figure}[htb!]
\centering
\includegraphics[scale=.5]{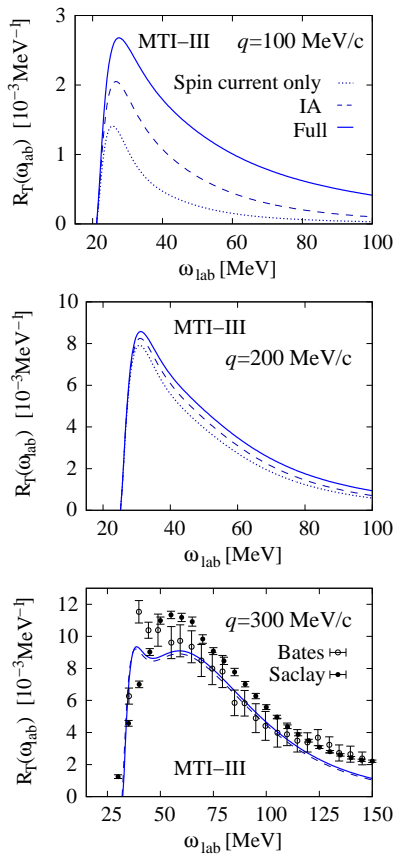}
\caption{(Color online)
Transverse response function for different momentum transfer values:
spin current only, IA and full calculation with the
inclusion of a consistent two-body current, in comparison with the
available experimental data from Bates~\cite{Dyt88} and
Saclay~\cite{Zgh94}. (For $q=300$ MeV/c the result
for the spin current only is already shown in Fig.~\ref{fig6}).} 
\label{fig9}
\end{figure}

\begin{table}
\caption{Maximal multipolarity $J_{max}$ considered in the calculation
shown in Fig.~\ref{fig9} for the one- and two-body currents and for
the different momentum transfer values.} 
\begin{ruledtabular}
\begin{tabular}{ccc}
$q$ [MeV/c]&  ${\boldsymbol j}_{(1)}({\boldsymbol q})$  &  ${\boldsymbol j}_{(2)}({\boldsymbol q})$  \\
\hline
100& 2 & 1\\
200 & 3   & 1\\
300 &  4  & 3\\
\end{tabular}
\end{ruledtabular}
\label{table}
\end{table}

In Fig.~\ref{fig9}, we finally present the transverse response
function for different momentum transfers in comparison to the
available experimental data. We compare the IA with the full
calculation in which we add the two-body currents. For $q=100$ and
200~MeV/c we also show the response function induced by the spin
current only. In Table \ref{table}, we recall explicitly the
multipoles considered in our final calculations of  $R_T(\omega,{\boldsymbol
q})$ for the different transverse current parts and for each momentum
transfer value. As previously shown in Fig.~\ref{fig6}, the spin
current strongly dominates the response function at a momentum transfer
$q=300$~MeV/c. A rather similar situation is found for $q=200$~MeV, as
depicted in Fig.~\ref{fig9}. On the other hand, one can see that for
the lower momentum transfer $q=100$~MeV/c the convection current presents a
stronger effect: for example, at $\omega=40$ MeV it is even half
of the total IA strength. By comparing the IA with the full
calculation, we can conclude that quite a strong effect of the MEC is
found at low momentum transfers like for $q=100$~MeV, where for example
at energies of 100 MeV they lead to an enhancement of the strength by
a factor of four with respect to the IA. At this momentum transfer,
the approximation we had introduced for the MEC is safely reliable. As
the momentum transfer increases, for example at $q=200$~MeV/c, the
enhancing effect of the MEC with respect to the IA result on $R_T$ is
reduced dramatically and it amounts to only $4\%$ in the peak region
and to $40\%$ for $\omega=100$ MeV. Unfortunately, to the best of our
knowledge, no experimental data have been measured in the quasi
elastic region at these low momentum transfers $q=100$ and
200~MeV/c. For the case of $q=300$ MeV/c where experimental data from
Bates \cite{Dyt88} and Saclay \cite{Zgh94} exist, we observe a tiny
effect of the two-body current with respect to IA result. The addition
of the MEC leads to an increase of $4\%$ and almost $10 \%$ at
$\omega=100$ and 150~MeV, respectively. Our full calculation agrees
rather well with the data of Bates in the energy range $55 \le \omega
\le 115$ MeV but is lower than the data of Saclay in the quasi-elastic
peak region and above. For example, at $\omega=60$ MeV our full result
for $q=300$~MeV/c is still lower than the measurement of Saclay. 

Here we would like to mention that the approximation introduced in the
MEC by neglecting the center of mass dependence, which is  estimated
to be of about $10\%$ to $20 \%$ in the peak and negligible in the tail
for $q=200$ to 300~MeV/c, does not affect our conclusion, due to
the fact that the overall MEC effect is small.

\section{Conclusions and outlook}
\label{Sec:Conclusions}

We have presented the first calculation of the inclusive longitudinal and
transverse response functions of $^4$He within the LIT and EIHH
methods with complete inclusion of the interaction in the final states
as given by the semirealistic potential MTI-III potential.
As in previous LIT calculations with the semirealistic TN potential \cite{EfL97} a
good overall agreement with available experimental data is found for
the longitudinal response function, though the MTI-III potential leads
in addition to the quasi-elastic peak to another peak close to
threshold, which is not clearly seen in the data of Bates~\cite{Dyt88}
and Saclay~\cite{Zgh94}, while, as mentioned, at $q< 100$ MeV/c a resonance
close to threshold has been observed in experiment \cite{Wal70}.

\begin{figure}[htb!]
\centering
\includegraphics[scale=.3]{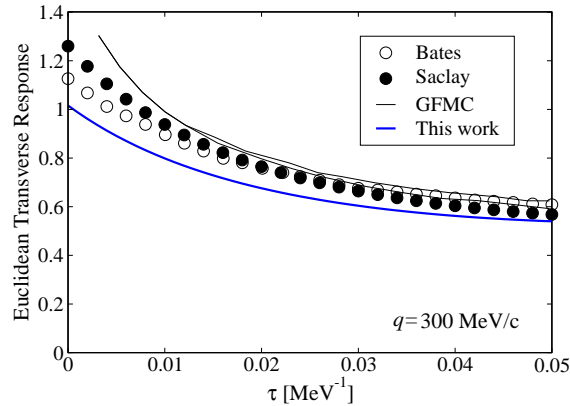}\\
\caption{(Color online) Euclidean transverse
response for momentum transfer $q=300$ MeV/c. Comparison of the GFMC
calculation \cite{CaS94} (band between thin lines) and the result of
this work (thick line) with the Laplace transform of experimental data
from Bates~\cite{Dyt88} and Saclay~\cite{Zgh94}.} 
\label{figure10}
\end{figure}

For the transverse response function, we have performed a calculation
in IA alone and then adding in a second step a two-body current
consistent with the MTI-III potential. Strong MEC effects are found at
low momentum transfer, $q=100$~MeV/c, where unfortunately no
experimental data are available. For the case of $q=300$~MeV/c, for
which data from Bates~\cite{Dyt88} and Saclay~\cite{Zgh94} exist, we have
shown that the IA result still misses some strength in the
quasi-elastic peak region. For the present MEC, consistent to the
semirealistic potential model, we have not found a strong two-body current
effect at $q=300$~MeV/c as obtained in the calculation of Carlson and
Schiavilla within the Laplace transform approach~\cite{CaS94}
 including $\pi$- and $\rho$-MEC. 
A direct comparison of the two calculations for $R_T(\omega,{\boldsymbol q})$ 
is not possible because 
the inversion of the Laplace transform suffers from large instabilities.
In Fig.~\ref{figure10}, we thus compare the Laplace transform of our
full transverse response function for $q=300$ MeV/c with the GFMC Euclidean
response, using as interaction the 
Argonne V18 potential \cite{wiringa:1995} with the addition of the
Urbana IX three-body force \cite{Urb9}. We adopt the definition of the
Euclidean response as presented in \cite{CaS92,CaS94} and we integrate
our theoretical curve up to the maximal value of energy transfer
($\omega_{max}=180$ MeV) measured in Saclay~\cite{Zgh94} for the
 considered momentum transfer case. Consistently with what is obtained in
 a direct comparison, our calculation for the MTI-III potential with
 the present MEC model leads  to a lower transverse Euclidean response
 with respect to the Laplace transform of experimental
 data. The Euclidean response obtained with the
 GFMC, where contributions of $\pi$- and $\rho$-exchange
 currents were considered, agrees better with data.
We have to conclude that the additional contribution of the present
MEC does not explain the missing strength of our model. The difference
to the GFMC calculation   could possibly be due to
the missing explicit pionic degrees of freedom in our potential model and
therefore also in the corresponding MEC. In particular, our MEC
model does not include any contact terms because the fictitious meson
exchanges comprise solely scalar mesons. It is well known, that such
contact terms as associated with pion exchange currents are quite
important. Moreover, the masses of the exchanged mesons of the
present potential model are considerably higher than the pion mass
making the exchange current shorter ranged and thus less important in
the range of momentum transfers considered in this work. Thus it might
not be surprising that we find a reduced effect of MEC.
On the other hand it is interesting to notice
that the IA result for MTI-III is closer to the data
than the corresponding one with AV18+UIX. This lets open the possibility
that for MTI-III the relative contribution of two-body currents
is indeed smaller than for a realistic interaction.
 Future studies
 should be extended to realistic potential models including
pion degrees of freedom.

\begin{acknowledgments}
The authors S.\ B. and H.\ A. would like to thank the Deutsche
Forschungsgemeinschaft for partial support (SFB 443).
This work was furthermore partially supported by the Israel Science
Foundation (Grant No.\ 361/05). 
\end{acknowledgments}

\appendix*
\renewcommand{\theequation}{A\arabic{equation}}
\section{Multipole expansion of the MEC}

Here, we will give a brief derivation of the multipole expansion of
a two-body current of the form
\beq
{\boldsymbol j}_{(2)}({\boldsymbol q},{\boldsymbol r},{\boldsymbol R})= e^{i{\boldsymbol q}\cdot {\boldsymbol R}}\, 
\widetilde {\boldsymbol j}_{(2)}({\boldsymbol q},{\boldsymbol r})
\eeq
consisting of an intrinsic part $\widetilde {\boldsymbol j}_{(2)}$, which depends
only on the relative two-body coordinate ${\boldsymbol r}$, and a two-body c.m.\ part,
depending on the two-body c.m.\ coordinate ${\boldsymbol R}$. The consistent
meson exchange current of (\ref{2b_curr_q}) has exactly this form. 

The basic quantities to evaluate are
\begin{equation}
J^{J}_{L,M}(q,{\boldsymbol r})=\frac{1}{4\pi}  \int d\hat{q}'  
e^{i{\boldsymbol q}\cdot {\boldsymbol R}} \boldsymbol{Y}^{J}_{L,M}(\hat{q}')
\cdot{\boldsymbol \nabla} \widetilde {\boldsymbol j}_{(2)}({\boldsymbol q}',{\boldsymbol r}) 
\,,\label{JJl}
\end{equation}
where $\boldsymbol{Y}^{J}_{L,M}$ denotes a vector spherical harmonics.
The quantities in (\ref{JJl}) determine the longitudinal and
transverse electric and magnetic multipoles according to
\beqa
L^J_0&=&-\sum_L (-)^L \widehat L 
\left(\begin{array}{ccc}
L&1&J \\ 
0&0&0
\end{array}\right)J^{J}_{L,0}\,,\label{long}\\
T^{el,J}_{\mu}&=&-\sqrt{2}\sum_{L=J\pm 1} (-)^L \widehat L 
\left(\begin{array}{ccc}
L&1&J\\ 
0&1&-1
\end{array}\right)J^{J}_{L,\mu}\,,\label{trans_e}\\
T^{mag,J}_{\mu}&=&-\sqrt{2} (-)^J \widehat J
\left(\begin{array}{ccc}
J&1&J\\ 
0&1&-1
\end{array}\right)J^{J}_{J,\mu}=J^{J}_{J,\mu}\label{trans_m}\,,
\eeqa
for $|\mu|=1$. Here a spherical coordinate system ${\boldsymbol \epsilon}_\mu$
with $\mu\in\{0,\pm1\}$ has been chosen where ${\boldsymbol \epsilon}_0$ is
along ${\boldsymbol q}$. For more details the reader should consult
Ref.~\cite{Bac05}.  

In order to evaluate (\ref{JJl}) one first expands
$e^{i{\boldsymbol q}\cdot {\boldsymbol R}}$ and the intrinsic current
${\boldsymbol j}_{(2)} ({\boldsymbol q},{\boldsymbol r})$ into spherical
harmonics. The former one is given by the well known expression
\beq
e^{i{\boldsymbol q}\cdot {\boldsymbol R}}=4\pi\sum_{J_R}(-)^{J_R}\hat J_R 
\Big[C^{J_R}(q,{\boldsymbol R})\times Y^{J_R}(\hat q)\Big]^0\,,
\eeq
where we have introduced the c.m.\ Coulomb multipoles
\beq
C^{J_R}(q,{\boldsymbol R})=i^{J_R}j_{J_R}(qR)Y^{J_R}(\hat R)\,,
\eeq
with spherical Bessel functions $j_l(x)$.
The corresponding expansion for the intrinsic current is given by
\beq
{\boldsymbol j}_{(2)} ({\boldsymbol q},{\boldsymbol r})=4\pi \sum_{J_rm_r L_r}
\widetilde J^{J_r}_{L_r,m_r}(q,{\boldsymbol r})
\boldsymbol{Y}^{J_r}_{L_r,m_r}(\hat{q})^*\,,
\eeq
where
\begin{equation}
\widetilde J^{J_r}_{L_r,m_r}(q,{\boldsymbol r})=\frac{1}{4\pi}  \int d\hat{q}'  
 \boldsymbol{Y}^{J_r}_{L_r,m_r}(\hat{q}')
\cdot{\boldsymbol \nabla} \widetilde {\boldsymbol j}_{(2)}({\boldsymbol q}',{\boldsymbol r})\,.
\end{equation}
Inserting these
expansions into (\ref{JJl}), then, after recoupling,
the angular integration can be done analytically, and one finds finally 
\beq
J^{J}_{L,\mu}(q,{\boldsymbol r},{\boldsymbol R})=(-)^{J+1}\sqrt{4\pi}\widehat J\widehat L
\sum_{J_rJ_rL_r}\widehat J_R\widehat L_r
\left(\begin{array}{ccc}
J_R&L_r&L\\ 
0&0&0
\end{array}\right)
\left\{\begin{array}{ccc}
J_R&J_r&J\\ 
1&L&L_r
\end{array}\right\}
\Big[C^{J_R}(q,{\boldsymbol R})\times \widetilde J^{J_r}_{L_r}(q,{\boldsymbol r})
\Big]^J_\mu\,.\label{mec_multipole}
\eeq
With the help of angular momentum algebra, the longitudinal and 
transverse multipoles are then obtained from
(\ref{long}) through (\ref{trans_m}), i.e.
\beqa
L^J_0(q,{\boldsymbol r},{\boldsymbol R})&=&(-)^J\frac{\sqrt{4\pi}}{\widehat J}
\sum_{J_RJ_r} {\widehat J}_R^2 {\widehat J}_r
\left(\begin{array}{ccc}
J_R&J_r&J\\ 
0&0&0
\end{array}\right)
\Big[C^{J_R}(q,{\boldsymbol R})\times \widetilde L^{J_r}(q,{\boldsymbol r})
\Big]^J_0\,,\label{mec_long}\\
T^{el/mag,J}_\mu(q,{\boldsymbol r},{\boldsymbol R})&=&(-)^{J+1}\frac{\sqrt{\pi}}{\widehat J}
\sum_{J_RJ_r} {\widehat J}_R^2 {\widehat J}_r
\left(\begin{array}{ccc}
J_R&J_r&J\\ 
0&-1&1
\end{array}\right)\Big(1\pm(-)^{J_r+J_r+J}\Big)\nonumber\\
&&\Big[C^{J_R}(q,{\boldsymbol R})\times \widetilde T^{el/mag,J_r}(q,{\boldsymbol r})
\Big]^J_\mu\,.\label{mec_trans}
\eeqa
Thus a total longitudinal or transverse multipole is given by a sum of
all possible couplings of corresponding intrinsic and c.m.\ charge
multipoles as allowed by parity and angular momentum coupling rules to
form a rank-$J$ tensor. 

With respect to the multipole expansion of the consistent MEC of
(\ref{2b_curr_q}) we start from the expansion ${\boldsymbol
\nabla}I_m$, which is given in~\cite{FaA76} 
\beq
{\boldsymbol \nabla}I_m({\boldsymbol q},{\boldsymbol r})=4\pi \sum_{J_r\mu L_r = 
{\rm even}}\widetilde I^{J_r}_{L_r\mu}(q,{\boldsymbol r},m)
\boldsymbol{Y}^{J_r}_{L_r\mu}(\hat{q})^*
\label{exp_Im}
\eeq
with
\beq
\widetilde I^{J}_{L\mu}(q,{\boldsymbol r},m)= 4\pi (i)^{J-1} \hat{L}
\left(\begin{array}{ccc}
1&L&J\\ 
0&0&0
\end{array}\right)
 Y^{J}_{\mu}(\hat{r})~\Phi^1_{J,L}(q,r,m)\,,
\label{mec_J_jlmu}
\eeq
where the functions $\Phi^{\nu}_{\sigma, l}(q,r,m)$ are defined by
\beq
\Phi^{\nu}_{\sigma, \ell}(q,r,m)=\frac{1}{q^2}\int_0^\infty 
\frac{dp p^{\nu}}{z} j_{\sigma}(pr) Q_{\ell}(z),
\label{phi_f}
\eeq
with $z=(p^2+\frac{1}{4}q^2+m^2)/{pq}$ and 
\beq
Q_{\ell}(z)=\frac{1}{2}\int_{-1}^{1}dx \frac{P_{\ell}(z)}{z-x}\,.
\eeq
Here, $P_{\ell}(z)$ denotes a Legendre polynomial. Then one obtains for
the intrinsic part of the MEC in (\ref{2b_curr_q})
\beqa
\widetilde J^{J_r}_{L_r\mu}(q,{\boldsymbol r})&=&\frac{4}{\pi}
({\boldsymbol \tau}_1 \times{\boldsymbol \tau}_2)_3
(i)^{J_r-1} \widehat{L}_r
\left(\begin{array}{ccc}
1&L_r&J_r\\ 
0&0&0
\end{array}\right) Y^{J_r}_{\mu}(\hat{r})\nonumber\\
&&\times\Big[\alpha\Phi^1_{J_r,L_r}(q,r,m_1)
+\beta\Phi^1_{J_r,L_r}(q,r,m_2)
+{\boldsymbol \sigma}_1 \cdot {\boldsymbol \sigma}_2
\Big(\gamma\Phi^1_{J_r,L_r}(q,r,m_1)
+\delta\Phi^1_{J_r,L_r}(q,r,m_2)\Big)\Big]\,,
\eeqa
which, inserted into (\ref{mec_long}) and (\ref{mec_trans}) finally
yields the MEC multipoles. 
From the condition $L_r ={\rm even}$ follows in conjunction with the
3j-symbol in (\ref{mec_J_jlmu}), that $\widetilde J^{J_r}_{L_r}(q,{\boldsymbol r})$ 
vanishes for $L_r=J_r$ and consequently all intrinsic magnetic
and even electric multipoles vanish according to (\ref{trans_e}) and
(\ref{trans_m}). 

If one neglects the c.m.\
contributions, i.e.\ setting ${\boldsymbol R}=0$, then one has
\beq
J^{J}_{L\mu}(q,{\boldsymbol r},{\boldsymbol 0})=\widetilde J^{J}_{L\mu}(q,{\boldsymbol r})\,.
\eeq
This approximation is used in the present work.

\end{document}